\documentclass[fleqn,usenatbib]{mnras}

\usepackage{newtxtext}

\usepackage[T1]{fontenc}

\DeclareRobustCommand{\VAN}[3]{#2}
\let\VANthebibliography\thebibliography
\def\thebibliography{\DeclareRobustCommand{\VAN}[3]{##3}\VANthebibliography}

\usepackage{graphicx}	
\usepackage{amsmath}	
\usepackage{amssymb}	
\usepackage{scalerel,tikz}
\usetikzlibrary{svg.path}
\definecolor{orcidlogocol}{HTML}{A6CE39}
\tikzset{orcidlogo/.pic={
 \fill[orcidlogocol] svg{M256,128c0,70.7-57.3,128-128,128C57.3,256,0,198.7,0,128C0,57.3,57.3,0,128,0C198.7,0,256,57.3,256,128z};
 \fill[white] svg{M86.3,186.2H70.9V79.1h15.4v48.4V186.2z}
 svg{M108.9,79.1h41.6c39.6,0,57,28.3,57,53.6c0,27.5-21.5,53.6-56.8,53.6h-41.8V79.1z M124.3,172.4h24.5c34.9,0,42.9-26.5,42.9-39.7c0-21.5-13.7-39.7-43.7-39.7h-23.7V172.4z}
 svg{M88.7,56.8c0,5.5-4.5,10.1-10.1,10.1c-5.6,0-10.1-4.6-10.1-10.1c0-5.6,4.5-10.1,10.1-10.1C84.2,46.7,88.7,51.3,88.7,56.8z};
}}
\newcommand\orcidicon[1]{\href{https://orcid.org/#1}{\mbox{\scalerel*{
\begin{tikzpicture}[yscale=-1,transform shape]
\pic{orcidlogo};
\end{tikzpicture}
}{|}}}}

\newcommand{\Msun}{{~\rm M_\odot}}

\newcommand{\kpc}{~\rm kpc}

\newcommand{\eagle}{\textsc{eagle}}
\newcommand{\tng}{\textsc{tng100-1}}
\newcommand{\auriga}{\textsc{auriga}}
\newcommand{\apo}{\textsc{apostle}}

\newcommand{\reffig}[1]{Fig.~\ref{#1}}

\title[Short title, max. 45 characters]{The galaxy size to halo spin relation of disk galaxies in cosmological hydrodynamical simulations}

\author[Yang et al.]{
Hang Yang $^{1,2}$\orcidicon{0000-0003-3279-0134} \thanks{E-mail: hyang@nao.cas.cn}, 
Liang Gao $^{1,2,3}$,
Carlos S. Frenk $^{3}$,
Robert J. J. Grand $^{4,5,6}$, 
Qi Guo $^{1,2}$, \newauthor\ 
Shihong Liao $^{7}$,
Shi Shao$^{3}$
\\
$^{1}$Key Laboratory for Computational Astrophysics, National Astronomical Observatories, Chinese Academy of Sciences, Beijing 100101, China\\
$^{2}$University of Chinese Academy of Sciences, 19 A Yuquan Rd, Shijingshan District, Beijing 100049, China\\
$^{3}$Institute for Computational Cosmology, Department of Physics, Durham University, Science Laboratories, South Road, Durham DH1 3LE, UK\\
$^{4}$Max-Planck-Institut für Astrophysik, Karl-Schwarzschild-Str. 1, 85748 Garching, Germany\\
$^{5}$Instituto de Astrofísica de Canarias, Calle Vía Láctea, E-38205 La Laguna, Tenerife, Spain\\
$^{6}$Departamento de Astrofísica, Universidad de La Laguna, Av. del Astrofísico Francisco Sánchez, E-38206, La Laguna, Tenerife, Spain\\
$^{7}$Department of Physics, University of Helsinki, Gustaf Hällströmin katu 2, FI-00014 Helsinki, Finland\\
}

\date{Accepted XXX. Received YYY; in original form ZZZ}

\pubyear{2020}

\begin{document}
\label{firstpage}
\pagerange{\pageref{firstpage}--\pageref{lastpage}}
\maketitle

\begin{abstract}
In the standard disk galaxy formation model, the sizes of galactic disks are tightly related to the spin parameters $\lambda$ of their dark matter haloes. The model has been wildly adopted by various semi--analytic galaxy formation models which have been extremely successful to interpret a large body of observational data. However, the size-$\lambda$ correlation was rarely seen in most modern hydrodynamical simulations of galaxy formation. In this short paper, we make use of 4 sets of large hydrodynamical simulations to explore the size-spin parameter relation with a large sample of simulated disk galaxies and compare it with a popular disk galaxy formation model of Mo et al. (1998). Intriguingly, galactic sizes correlate with spin parameters of their dark matter haloes in the simulations developed by the IllustrisTNG collaborations, albeit the relation does not always agree with prediction of MMW98 model overall stellar mass range we examined. There is also a size-spin correlation for the Milky way analogies in the EAGLE simulations, while it is relatively weaker than that of the IllustrisTNG counterparts. For the dwarfs in the simulations from the EAGLE collaboration, there is NULL correlation. We conclude that either the detailed subgrid physics or hydrodynamics solvers account for the size-spin parameter relation, which will be explored in our future work.
\end{abstract}




\section{Introduction}

In the classic galaxy formation theory \citep[][]{1978MNRAS.183..341W, 1980MNRAS.193..189F, 1991ApJ...379...52W}, galaxies form in two-stages: dark matter collapse to form self-bound dark matter haloes due to gravitational instability; because of radiative cooling, baryons condense in centres of dark matter haloes to form gaseous disk as a consequence of angular momentum conservation. These cold gas later further fragment and form luminous galaxies when certain conditions are satisfied. 

In this framework, since the baryons and dark matter are expected to be initially well mixed and hence experience similar tidal torques \citep[][]{1969ApJ...155..393P, 1984ApJ...286...38W}, the galactic disk, which is a consequence of gas condensation, should have similar specific angular momentum as its dark matter halo, namely $j_{d} \sim j_{h}$. Here $j_d$ and $j_h$ are specific angular momentum of galaxy and halo respectively. The specific angular momentum of a dark matter halo $j_h$ is often characterized by a dimensionless spin parameter $\lambda$ \citep[][]{2001MNRAS.321..559B}, which can be written as
\begin{equation}
 \lambda=\frac{j_h}{\sqrt{2}V_{200}R_{200}}.
 \label{equ:spin}
\end{equation}
Where $R_{200}$ represents the virial radius of a dark matter halo within which the enclosed mean density is 200 times the critical density of the Universe and $M_{200}$ is the mass enclosed within $R_{200}$. $V_{200}$ is the virial velocity of the halo,  $V_{200}=\sqrt{GM_{200}/R_{200}}$. 

Assuming the angular momentum of the stellar disk is a fraction $f_j=j_d/j_h$ of the halo, \citet{1998MNRAS.295..319M} (hereafter MMW98) links the size of a disk galaxy $r_d$ and virial radius $R_{200}$ of its host dark matter halo with a form as \footnote{Compared with the original MMW98 model, we use a different definition for spin parameter $\lambda$. Hence, there is no $f_c$ factor here.}

\begin{equation}
  \frac{r_{1/2}}{R_{200}}= \frac{1.68}{\sqrt{2}}f_{j}f_R\lambda 
  \label{equ:MMW98} 
\end{equation}

Where the $f_R$ factor is introduced to account for the different rotation velocity curve of the galaxy due both to dark matter adiabatic contraction \citep[][]{1986ApJ...301...27B} and the self gravitational effects of the disk.

The angular momentum-based models (e.g. MMW98) have been successful to explain the observed distribution of disk scale lengths \citep[e.g.][]{2003MNRAS.343..978S, 2008ApJ...672..776S, 2013ApJ...764L..31K, 2017ApJ...838....6H, 2018ApJ...859....2L, 2020MNRAS.492.1671Z, 2020A&A...644A..76P}, and been widely used in various semi-analytic models \citep[e.g.][]{2000MNRAS.319..168C, 2003MNRAS.343...75H, 2006MNRAS.365...11C, 2007MNRAS.375....2D, 2008MNRAS.391..481S} by adopting both $f_j$ and $f_{R}$ to be units. Recent semi-analytic models have been improved by assuming the cooling gas carrying the same specific angular momentum as that of the host halo at each time step, which later is added to the stellar disk via star formation \citep[e.g.][]{2009MNRAS.396..141D, 2011MNRAS.413..101G}. These semi-analytic models, combined with N-body simulations, have been very successful to match a large body of observables, including galaxy size and morphological types, etc. In addition, this improvement makes the model predictions more in line with later studies that the angular momentum vector of the gas and dark matter are not necessary to be identical \citep[e.g.][]{2005ApJ...628...21S, 2009MNRAS.399L..64S, 2017MNRAS.470.2262L, 2018MNRAS.475..232P, 2019MNRAS.489.3609I}.

The formation of disk galaxy has also been extensively investigated with hydrodynamical simulations \citep[e.g.][]{1991ApJ...377..365K, 1994MNRAS.267..401N, 1999ApJ...513..555S}. Until recently, with significant progress in sub-grid physics models, in particularly the feedback model, many modern hydrodynamical galaxy formation simulations are able to reproduce galaxies with different morphological types \citep[e.g.][]{2012MNRAS.423.1726S, 2014MNRAS.444.1453D, 2014MNRAS.442.2304H, 2014MNRAS.444.1518V, 2015MNRAS.446..521S, 2015ApJ...812...29T}. However, studies based on some of these modern hydro-dynamical simulations, suggested that, while sizes of the simulated galaxies are statistically proportional to the virial radius of their host dark matter haloes, there are no correlations between halo spin parameters $\lambda$ \citep[][]{2019MNRAS.488.4801J, 2015MNRAS.454...83W, 2014MNRAS.442.1545C, 2015MNRAS.450.2327Z}. This result challenges the classical theory and many existing semi-analytical models. On the contrary,  \citet{2017MNRAS.471L..11D} found a weak correlation between galaxy size and host halo spin parameter in the \eagle{} simulation. \citet{2019MNRAS.490.5182L} found that there is a strong correlation between sizes and host halo spin parameters $\lambda$ for field dwarf galaxies in the \auriga{} simulation. 

In this paper, we use 4 sets of high-resolution hydrodynamical simulations, to explore the relation between the sizes and spin parameters of host dark matter haloes of a large sample of simulated disk galaxies, and explicitly compare them with predictions from the MMW98. The paper is organized as follows. In Section 2, we briefly introduce the numerical simulations and methodology used in this study. The main results are presented in Section 3, and conclusions are drawn in Section 4.

\section{The Simulations and Methodology}
\label{sec:Simulation}

The numerical simulations used in this paper comprise 4 suits of large hydrodynamical galaxy formation simulations, the \textsc{IllustrisTNG} \citep[][]{2016MNRAS.457.1931S}, \auriga{} \citep[][]{2017MNRAS.467..179G}, \eagle{} \citep[][]{2015MNRAS.450.1937C}, and \apo{}-L2 projects \citep[][]{2016MNRAS.457.1931S}. The former two simulations are performed with the same hydrodynamical scheme of the \texttt{AREPO} \citep[][]{2010MNRAS.401..791S} code and with similar subgrid physics models developed by the Illustris collaborations, the latter two are run with the improved smoothed particle hydrodynamics (SPH) and identical subgrid physics model developed by the EAGLE collaborations. \citep[][]{2015MNRAS.450.1937C}

\subsection{The simulations}
The \textsc{IllustrisTNG} project \citep[][]{2018MNRAS.475..648P, 2018MNRAS.475..676S, 2018MNRAS.475..624N, 2018MNRAS.480.5113M, 2018MNRAS.477.1206N} is a suite of cosmological magneto-hydrodynamic simulations, which was performed with the magneto-hydrodynamic moving mesh code \texttt{AREPO} \citep[][]{2010MNRAS.401..791S}. The \textsc{IllustrisTNG} project assume $\Omega_m=0.3089$, $\Omega_b=0.0486$,
$\Omega_{\Lambda}=0.6911$,$h=0.6774$, $n_s=0.9667$ and $\sigma_8=0.8159$
\citep[][]{2014A&A...571A..16P}. In this work, we use the \tng{} project with a box size about 110Mpc. We refer the reader for the detailed galaxy formation models of the \textsc{IllustrisTNG} simulations to \citet{2017MNRAS.465.3291W} and \citet{2018MNRAS.473.4077P}.

The \auriga{} project \citep[][]{2017MNRAS.467..179G} comprises a suite of 30 zoom-in cosmological simulations of Milky Way-sizes haloes and their surroundings. The parent haloes in \auriga{} were selected from a dark matter only simulation \eagle{} (L100N1504) \citep[][]{2015MNRAS.446..521S}. Similar to the \tng{}, the \auriga{} projects were performed with the  magneto-hydrodynamic moving mesh code \texttt{AREPO} \citep[][]{2010MNRAS.401..791S}, but assume slightly different Cosmological parameters, $\Omega_m=0.307$, $\Omega_b=0.048$, $\Omega_{\Lambda}=0.693$, $h=0.6777$,  $n_s=0.9611$ and $\sigma_8=0.829$ \citep[][]{2014A&A...571A..16P}. 

The \eagle{} project \citep[][]{2015MNRAS.446..521S, 2015MNRAS.450.1937C} is a suite of cosmological hydrodynamic simulations, which were performed with a version of the N-body Tree-PM smoothed particle hydrodynamics (SPH) code \texttt{GADGET-3} by \citet{2005MNRAS.361..776S}. The cosmological parameters adopted in the \eagle{} project are $\Omega_m=0.307$, $\Omega_b=0.048$, $\Omega_{\Lambda}=0.693$,$h=0.6777$, $n_s=0.9611$ and $\sigma_8=0.829$ \citep[][]{2014A&A...571A..16P}. In this work, we use the Ref-L0100N1504 run which has a volume of $(100Mpc)^3$. 

The \apo{} project \citep[][]{2016MNRAS.457.1931S, 2016MNRAS.457..844F} performed a suite of cosmological hydrodynamic zoom-in simulations of 12 volumes selected to match the kinematics of Local Group. High-resolution regions of the \apo{} were selected from dark matter only simulation \textsc{DOVE} which evolved a cosmological volume of ($100Mpc)^3$. The \apo{} project was performed with the same code \texttt{GADGET-3} as \eagle{}, and run with three different resolutions: low(L1), medium(L2), and high(L3). Since only two volumes have been run at high-resolution (L3) in \apo{}, we use the medium-resolution (L2) data in this work. The cosmological parameters in \apo{} simulation adopt the result of WMAP-7, namely $\Omega_ m=0.272$, $\Omega_b=0.0455$, $\Omega_{\Lambda}=0.728$,$h=0.704$, $n_s=0.967$ and $\sigma_8=0.81$ \citep[][]{2011ApJS..192...18K}.

The table~\ref{tab:simulation} summarize the typical individual particle mass and softening length for all the above simulations.

\begin{table}
	\centering
	\caption{Numerical parameters of the simulations used in this study. The columns shows: (1) softening length (2) baryonic particles mass; (3) dark matter particles mass.}
	\label{tab:simulation}
	\begin{tabular}{lcccc} 
		\hline
		 & $\epsilon[pc]$ & $m_{b}[M_{\odot}]$ & $m_{DM}[M_{\odot}]$ \\
		\hline
		AURIGA & 369 & $5\times{10^4}$ & $3\times{10^5}$ \\
		TNG100-1 & 740 & $1.4\times{10^6}$ & $7.5\times{10^6}$ \\
		APOSTLE-L2 & 216 & $1.2\times{10^5}$ & 5.8$\times{10^5}$ \\
		EAGLE(RefL0100N1504) & 700 & $1.8\times{10^6}$ & $9.7\times{10^6}$ \\
		\hline
	\end{tabular}
\end{table}

In all the above simulations, dark matter haloes are identified with  friends-of-friends (FOF) algorithm \citep[][]{1985ApJ...292..371D} and subhaloes are subsequently identified with the SUBFIND algorithm \citep[][]{2001MNRAS.328..726S, 2009MNRAS.399..497D}.

\subsection{Determination of galaxy morphology and halo spin parameter}

In order to reliably measure the sizes of the simulated galaxies, we include all central galaxies containing at least $250$ \footnote{We have also selected galaxies that have at least $1000$ stellar particles \citep[][]{2019MNRAS.487.5416T} and found the result remains are qualitatively similar.} stellar particles and host halo mass satisfy $\log{(M_{200}/\Msun)}<12.3$. These galaxies span almost four orders of magnitude in stellar mass and reside in a variety of environments. We further discard the galaxies contaminated by low-resolution particles in the zoom-in \auriga{} and \apo{} simulations. We have excluded all the satellites but only use the central galaxies in this study. The final galaxy sample contains 19315 galaxies from the \tng{}, 282 galaxies from the  \auriga{}, 12327 galaxies from the \eagle{}, and 408 galaxies from the \apo{}.

We define the morphology of each galaxy of the above galaxy sample by introducing the $\kappa$ parameter defined as the ratio of rotational kinetic energy $K_{rot}$ to total kinetic energy $K$ for a galaxy \citep[][]{2012MNRAS.423.1544S}, written as
\begin{equation}
 \kappa= \frac{K_{rot}}{K}=\frac{\sum_i{1/2}m_i\{(\hat{\boldsymbol{L}}\times{\hat{\boldsymbol{r}}_i})
 \cdot{\boldsymbol{v}_i}\}^2}
{\sum_i{1/2}m_i{v_i^2}},
\label{equ:Kappa}
\end{equation}

Where $\hat{\boldsymbol{L}}$ is the unity total angular momentum vector of stellar components. The $m_i$, $r_i$ and $v_i$ are mass, position vector to centre, velocity vector to centre for stellar particle $i$, respectively.

\begin{figure}
    \centering
	\includegraphics[width=\linewidth]{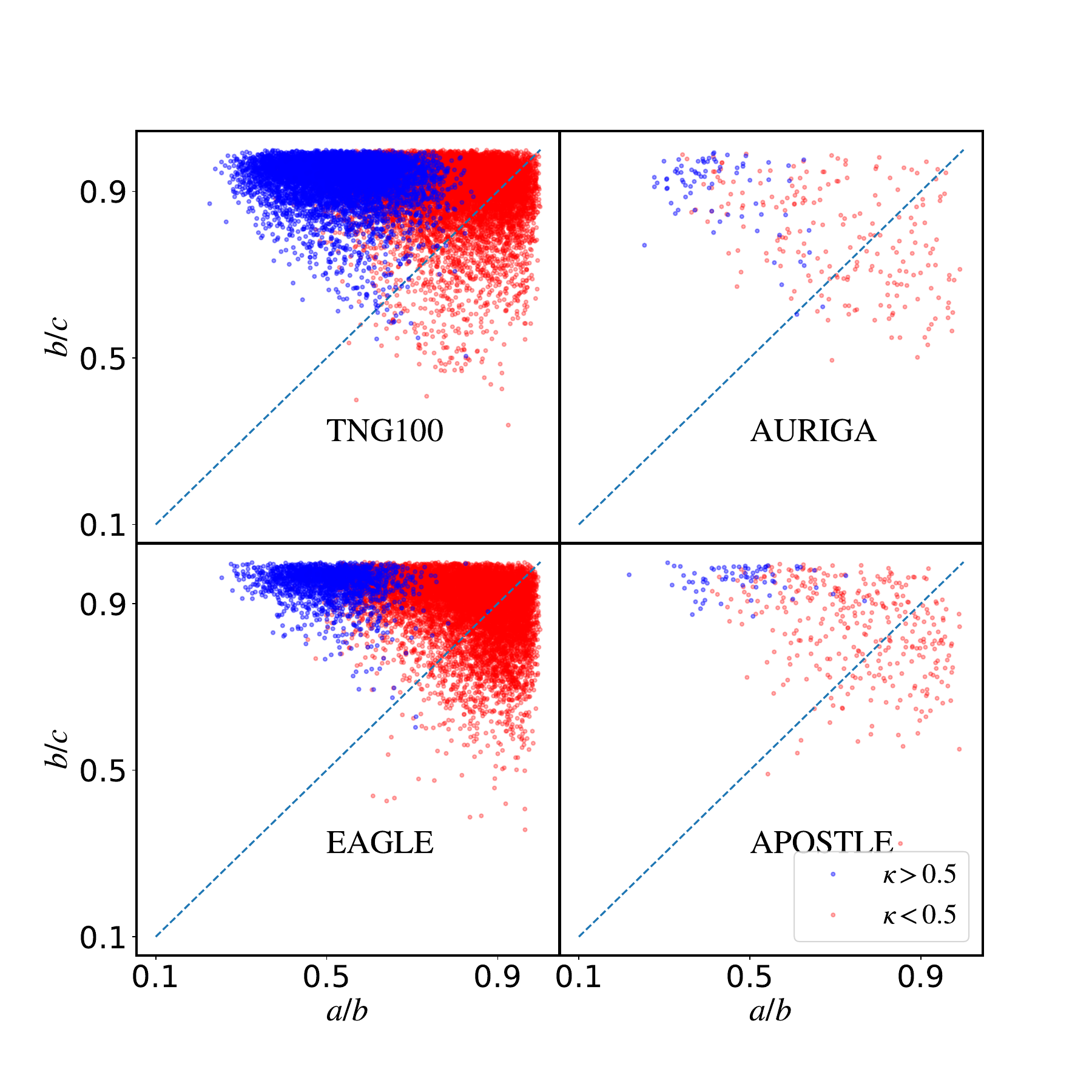}
    \caption{Axial ratio $b/c$ verus $a/b$ of each simulated galaxy in our sample. The blue dots represent disk-like galaxies, the red dots represent spheroidal-like galaxies, and the blue dash lines in each panel indicate $b/c=a/b$ ($b/c>a/b$ for oblate, $b/c<a/b$ for prolate). }
    \label{fig:morphology}
\end{figure}

For each galaxy in our sample, we calculate its $\kappa$ parameter with all star particles within 2 times of its half-stellar-mass radius, $2r_{1/2}$. A galaxy is classified as a disk (or spheroidal) galaxy if its $\kappa > (<)0.5$. Note, $\kappa$ is a definition of morphology according to kinematics and correlates strongly with the axial ratios of a galaxy. In Figure~\ref{fig:morphology} we show $b/c$ versus $a/b$ of our galaxy sample in different simulations. Here the axial ratios of galaxies are obtained by diagonalizing the inertia tensor matrix,

\begin{equation}
    I_{\alpha\beta}=\Sigma_i{m_i}(x_{i,\alpha}-x_{c,\alpha})
    (x_{i,\beta}-x_{c,\beta})
\label{equ:ITM}
\end{equation}
Where $x_i$ is the spatial position for particle $i$ and $x_c$ is the position with the minimal gravitational potential for the galaxy. We calculate the inertia matrix by the stellar components within twice the stellar half mass radius, $a<b<c$ are eigenvalues of the inertia tensor matrix $I_{\alpha\beta}$.

The disk galaxies are shown as blue dots and spheroidal galaxies are shown as red ones. As can be seen clearly that the classification of galaxy morphology type with $\kappa=0.5$ is reasonable. We also tested other critical values for $\kappa$(from $0.4$ to $0.5$), and it had little effect on the final correlation results. We summarize the number of various samples in the table~\ref{tab:sampleSize}. It is worth noting that our kinematic morphological classification differs from the standard photometry-based method used in observation, with the two showing a moderate correlation with considerable scatter \citep[e.g.][]{2003ApJ...597...21A, 2010MNRAS.407L..41S}.

\begin{table}
	\centering
	\caption{Sample number of the simulations used in this study. The columns shows: (1) total sample number (2) disk galaxies number(${\kappa}>0.5$); (3) spheroidal galaxies number(${\kappa}<0.5$).}
	\label{tab:sampleSize}
	\begin{tabular}{lcccc} 
		\hline
		 & all & disk & spheroidal \\
		\hline
		AURIGA & 282 & 77 & 205 \\
		TNG100-1 & 19315 & 6615 & 12700 \\
		APOSTLE-L2 & 408 & 71 & 337 \\
		EAGLE(RefL0100N1504) & 12327 & 1831 & 10496\\
		\hline
	\end{tabular}
\end{table}

For each halo in our sample, we calculate its dimensionless spin parameter $\lambda$ using the formula given by \citep[][]{2001MNRAS.321..559B}.
\begin{equation}
 \lambda=\frac{j_h(<R_{200})}{\sqrt{2}V_{200}R_{200}}
\end{equation}

\begin{figure}
	\includegraphics[width=\linewidth]{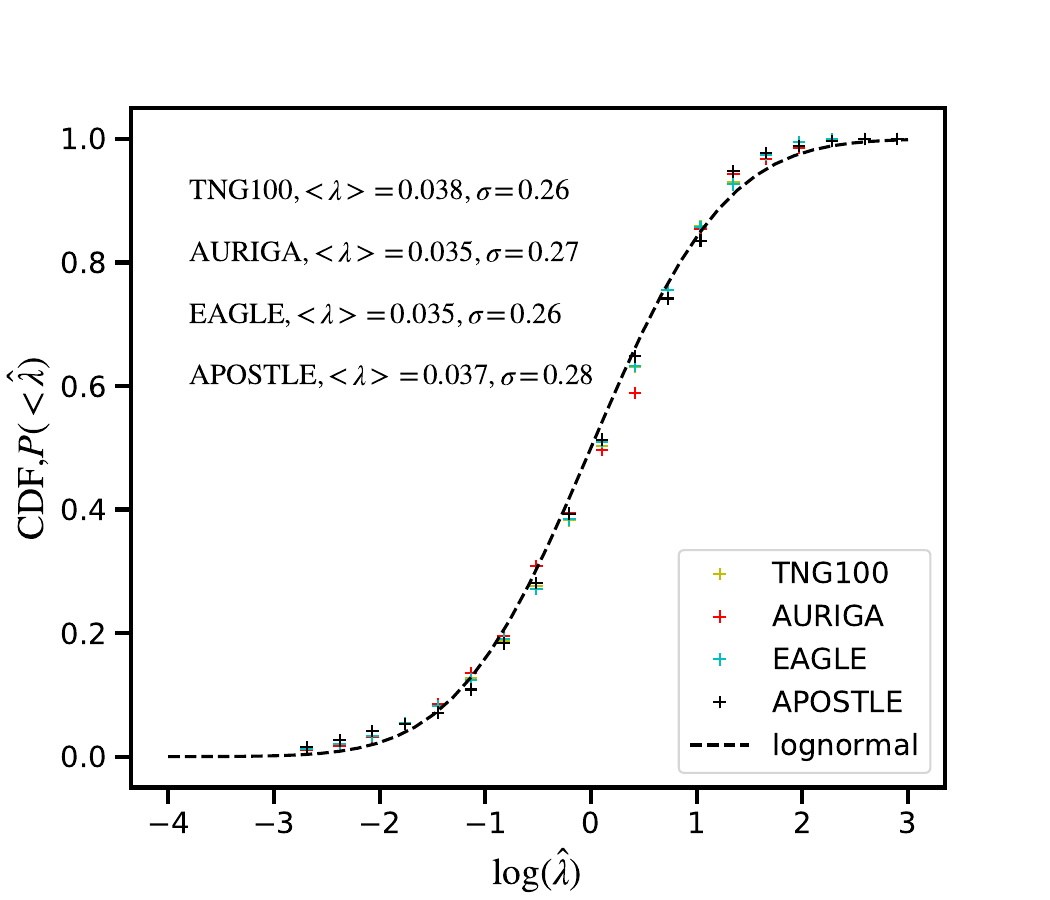}
	\centering
    \caption{Cumulative distribution of normalizational spin parameter $\hat{\lambda}=(\log_{10}{\lambda}-{<\log_{10}{\lambda}>})/\sigma$ of all dark matter halo samples in different simulations. Results from different simulations are shown in different colors as indicated on the label. The dashed lines show a standard single log-normal distribution. Mean values and standard deviation of spin parameters of different simulations are also shown in the figure.}
    \label{fig:spin}
\end{figure}

Previous works have shown that the distribution of halo spin parameter $\lambda$ is independent of halo mass and follow a log-normal distribution with the mean value $\left<\lambda\right>\sim{0.03-0.04}$ and standard deviation $\sigma_{\log_{10}\lambda}\sim{0.2-0.3}$ \citep[][]{2007MNRAS.378...55M, 2007MNRAS.376..215B, 2019MNRAS.488.4801J}. In figure~\ref{fig:spin}, we show the cumulative halo spin parameter distributions of our halo sample of all simulations used in this study without morphology cut, results for different simulations are distinguished with different colors as indicated in the label. Mean values and standard deviation of spin parameters of different simulations are also shown on the label. Clearly, these results are consistent with each other and with previous works. 

\section{Results}
\subsection{Galaxy size-stellar mass relations}

\begin{figure*}
	\includegraphics[width=0.7\textwidth]{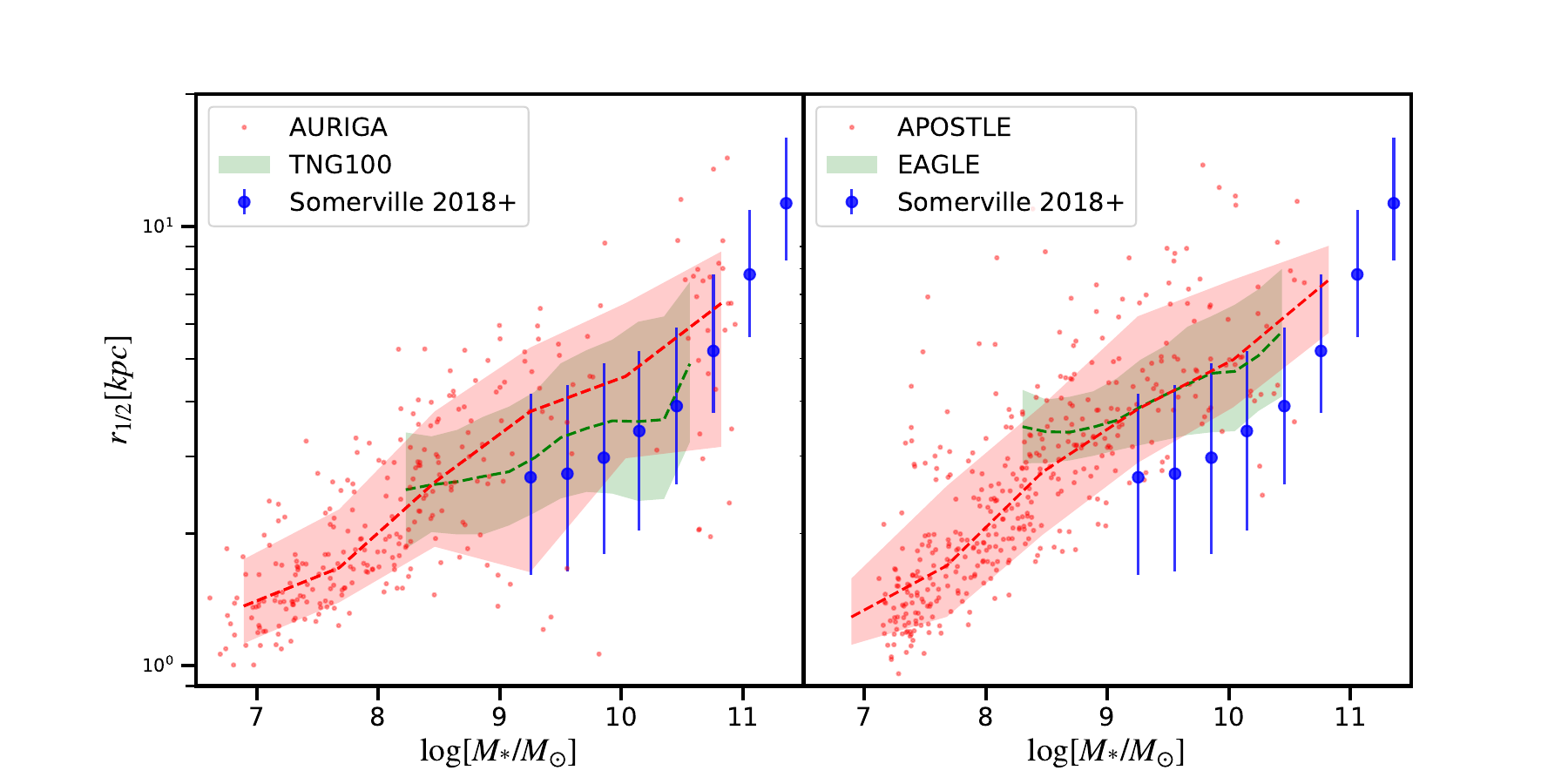}
	\centering
    \caption{Galaxy size-mass relation in different simulations (\textit{left panel}: \auriga{} and \tng{} simulations; \textit{right panel}: \apo{} and \eagle{} simulations) at $z=0$. Shaded regions show 16th to 84th percentiles of the result in each stellar mass bin. The thick blue points show the observed relation by \citet{2018MNRAS.473.2714S}, and error bars show 1$\sigma$ scatter.}
    \label{fig:size}
\end{figure*}

\begin{figure*}
	\includegraphics[width=0.75\textwidth]{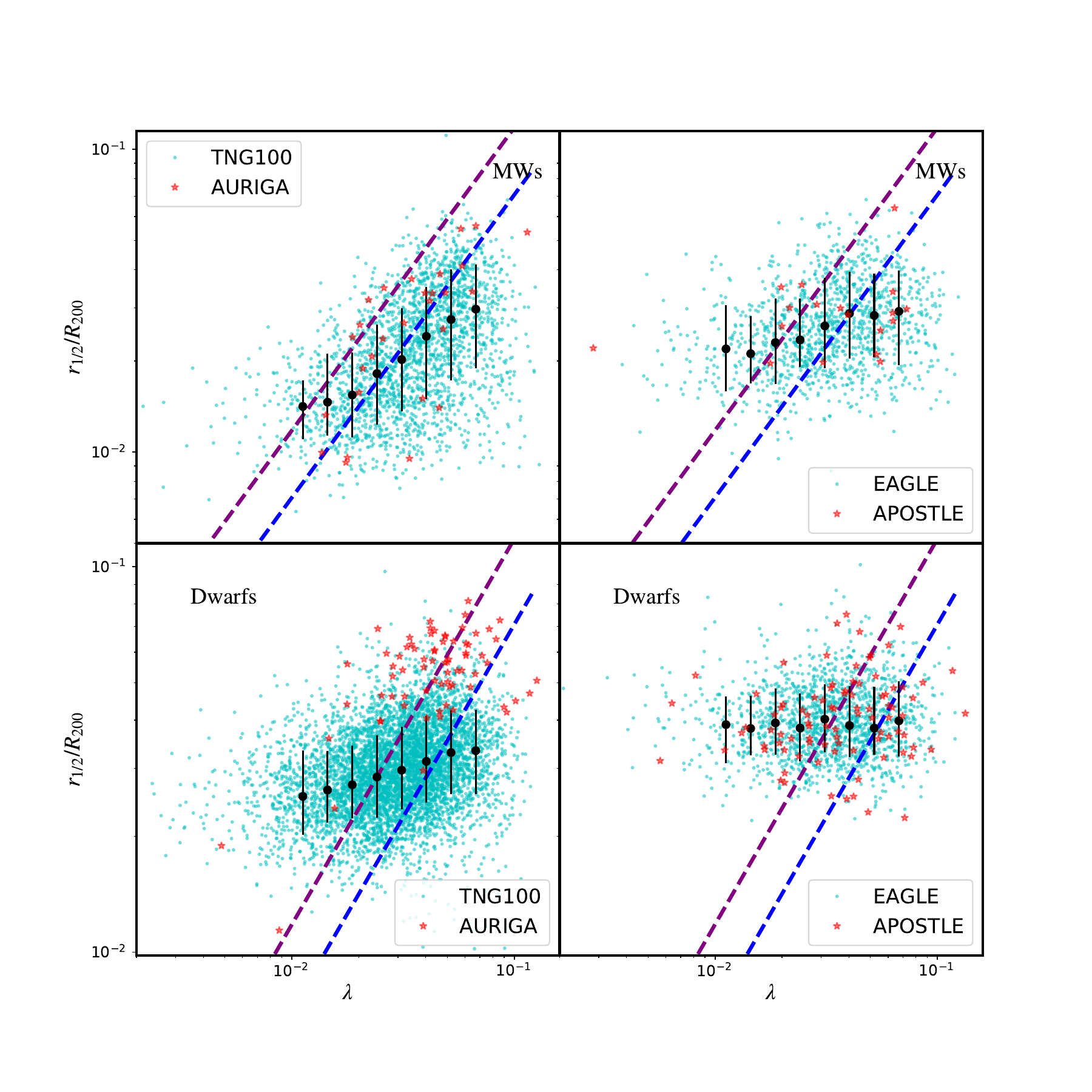}
	\centering
    \caption{Galaxy $r_{1/2}/R_{200}$ vensus host halo spin $\lambda$ of disk galaxies in our galaxy sample. Black dots show median values of the TNG100 (left panels) or Eagle (right panels) disk galaxies, respectively. Error bars show
    16th and 84th percentiles in each $\lambda$ bin. The dashed line display predictions given by the MMW98 model assuming $f_j=1.0$ (purple) and $f_j=0.5$ (blue) with $f_R=1$, respectively.}
    \label{fig:res3}
\end{figure*}

Compared with observations, we adopt a different morphology definition method ($\kappa$) here. In Figure \ref{fig:size} we present the size-stellar mass relation of all simulated galaxies at $z=0$ without morphology cutting, and then fairly compare the results with observations. The results for the \auriga{} and \tng{} are shown in the left panel and those for the \apo{} and \eagle{} are shown in the right panel, results for different simulations are distinguished with different colors as shown in the label. The shaded areas show $1\sigma$ scatters of each simulation, and the median values are shown with dashed lines. Here the size of a simulated galaxy is defined as the half-mass radius within which the enclosed stellar mass is half of the whole galaxy. In all simulations, galaxy sizes increase with increasing galaxy stellar masses with $r_{1/2}\sim1\kpc$ for low-mass dwarf galaxies, and $r_{1/2}\sim5\kpc$ for MW-mass galaxies. We find overall a good agreement between the full-box simulations (\tng{} and \eagle{}) and their zoom-in counterparts (\auriga{} and \apo{}) at the mass where the two simulations overlap. For masses $M_{\star}\leq10^{10.5}\Msun$, the size-mass relation of the galaxies in \tng{} and \eagle{} is almost flat, while the slope of the zoom-in galaxies in \apo{} or \auriga{} is much steeper, the small difference in the slope is due to the effect of simulation resolution on the size of the galaxies as studied in prior literature \citep[e.g.][]{2019MNRAS.488L.123L, 2018MNRAS.475..648P}.

We then compare the size-mass relations of simulated galaxies with an observational study by \citet{2018MNRAS.473.2714S} (blue dots with error bars ). The conversion of the observational projected semimajor half light radius $r_{e,2d}$ into $r_{1/2}$ involves two factors, 
\begin{equation}
    r_{e,2d} = f_pf_kr_{1/2}
\label{equ:retor1/2}
\end{equation}
The projection correction factor $f_p=1 (0.68)$ and the light to mass weighting factor $f_k=1.2 (1.15)$ for disk (spheroidal) galaxies in \citet{2018MNRAS.473.2714S}. The observational results clearly also show an increasing slope for galaxies with stellar mass. At masses, $M_{\star}\geq10^{9.5}\Msun$, the agreement between \tng{} and observations is rather good (see the left panel in \reffig{fig:size}), consistent with the study by \citet{2018MNRAS.474.3976G}. \auriga{} uses the same hydrodynamics solver and similar subgrid physics with \tng{}. Thus, the sizes of \auriga{} central galaxies that have stellar mass, $M_{\star}\geq10^{10.5}\Msun$, also agree with the observations.

For galaxies in \eagle{} and \apo{} (right panel in \reffig{fig:size}), the sizes are systematically about $0.17$ dex larger than the observed values, whereas \citet{2017MNRAS.465..722F} who uses the same \eagle{} data found a better agreement with observation than our result. The discrepancy could be due to the fact that \citet{2017MNRAS.465..722F} selected galaxies in a redshift bin of $\Delta z = 0.5$ while we only make use of galaxies at $z=0$, and the inclusion of high-redshift galaxies could reduce the median of the galaxy sizes. A secondary effect is a different definition of disk galaxies. The final effect is that our definition of size $r_{1/2}$ is typically larger than if we take into account the stellar particles within a specific spherical aperture. \citet{2017MNRAS.465..722F} adopted an aperture of radius, $r = 100 $\kpc{}, to exclude the stellar particles that belong to the galaxy by the subhalo but are located far out. However, the aperture measurements only affect very high-mass galaxies with $M_{\star}\geq10^{10.5}\Msun$ which is larger than most of our selected galaxies. In addition, the results for passive and active galaxies was separated when Furlong compared \eagle{} with observations. 

\subsection{Galactic size-host halo spin parameter relations}

In Figure~\ref{fig:res3}, we show $r_{1/2}/R_{200}$ of each simulated galaxy ($\kappa>0.5$) of our sample versus spin parameter $\lambda$ of its host halo. The upper panels show results for Milky way sized galaxies and the bottom show results for dwarfs, results from different simulations are shown with different symbols as indicated in the label. Here we follow \citep[e.g.][]{Wang2015, 2020SCPMA..6309801W} to define Milky way sized galaxies as the galaxies whose halo masses are in the range, $M_{\rm 200} \in [0.5, 2] \times 10^{12}\Msun$, and dwarfs as those with halo masses ranging from $M_{\rm200}{\leq}1.5{\times}10^{11}\Msun$. The thick black dots indicate the median values of each $\lambda$ bin in the \tng{} and \eagle{} simulations. Error bars show the 16th to 84th percentile of the size ratios. There is a strong correlation between the $r_{1/2}/R_{200}$ and $\lambda$ for $\lambda \geq 0.01$ in the \tng{} and \auriga{}, while there is very weak or no correlation below the value. Interestingly, the $\lambda$-size relation is much weaker in the \eagle{} and \apo{} simulations. Table~\ref{tab:res2} summarizes the values of the Spearman correlation coefficient $\rho_s$ of all simulations. 

The purple and blue dashed lines in all panels show predictions of \citet{1998MNRAS.295..319M} models, assuming $f_j=1$ and $f_j=0.5$, respectively. Interestingly, MMW98 model with $f_j=0.5$ agrees reasonably with \tng{} and \auriga{} for the MWs samples, while the agreement is worse for dwarfs, even though the \auriga{} dwarfs seems to agree with MMW98 with $f_j=1$. It is noticeable that the $r_{1/2}/R_{200}$ of the \auriga{} disk dwarfs tends to be overall larger than those of the \tng{}. The reason may be due to that the size of \auriga{} sample is smaller than the \tng{}. For dwarf-sized dark matter haloes in the \eagle{} and \apo{}, the size--$\lambda$ relation is almost null. We have to note that the slope and amplitude of the size--$\lambda$ relation may also rely on the halo concentration and angular momentum retention factor $f_j$ (see appendix A and B respectively). Galaxies with lower concentration have larger size in MW-mass sample and galaxies with higher $f_j$ have larger size in all sample.

\begin{table}
	\centering
	\caption{Spearman correlation coefficients $\rho$ for $\lambda-r_{1/2}/R_{200}$ relation of simulated disk galaxies. Error bars are estimated with Fisher transformation method assuming $95\%$ confidence.}
	
	\label{tab:res2}
	\begin{tabular}{lccccc} 
		\hline
		 & MW-like & Dwarf  \\
		\hline
		$\rho$(TNG100\&Auriga) & 0.50 $\pm$ 0.05 & 0.38  $\pm$ 0.05 \\
		$\rho$(Eagle\&Apostle) & 0.32 $\pm$ 0.07 & 0.02  $\pm$ 0.09 \\
		\hline  
	\end{tabular}
\end{table}

\section{Discussions and conclusions}

In the classic picture of the disk galaxy formation model, the sizes of disk galaxies are tightly related to the spin parameter of their dark matter haloes. In this short paper, we make use of 4 sets of modern hydrodynamic simulations of galaxy formation to examine this scenario, and compare results of the simulations with a popular disk galaxy formation model of MMW98. Our results can be summarized as follows.

Galaxy size--stellar mass relations in the IllustrisTNG and \eagle{} simulations agree reasonably with observational results, while the agreement is better in the Illustris \textsc{TNG100}; the relation seems convergent in both sets we used, of the Illustris and \eagle{}. For the simulated disk galaxies selected with $\kappa$, there are moderate correlation between size ratio, $r_{1/2}/R_{200}$, and spin parameter, $\lambda$, of dark matter halo in the Illustris family runs, \tng{} and \auriga{}, while the correlation is weak or null in the \eagle{} and \apo{} simulations. The scatter of the $r_{1/2}/R_{200}$--$\lambda$ relation could be due to the variety in the halo concentration and retention factor $f_j$. The spearman correlation coefficient is $0.50$ ($0.38$) for the Milky Way sized disk galaxies in the Illustris family(\eagle{} family), and $0.32$ ($0.02$) for the disk dwarfs in the Illustris family(\eagle{} family).  The size--spin parameter relation of the simulated MWs in the \tng{} and \auriga{} simulations agree well with MMW98 model by assuming $f_j=0.5$, but not for the dwarfs which have different logarithmic slopes from the prediction of the same model.

Intriguingly, on the classic disk formation model, the results from the Illustris and \eagle{} family runs are nearly opposite. While the Illustris runs are in qualitatively support of the model, the \eagle{} runs, along with an existing study from NIHAO simulation \citep[][]{2019MNRAS.488.4801J}, are largely against it. As the hydrodynamic solvers and detailed subgrid physics implemented in different galaxy formation models discussed here are quite different, it is unclear what is the dominant factor to set up the halo spin and stellar disk size relation seen in the Illustris runs. A pioneering work has shown that \eagle{} and \auriga{} exhibit different gas properties in the Milky Way-mass galaxies, baryon cycle is almost closed in the \auriga{} main galaxy while less baryons reside within the halo in \eagle{} \citep[][]{2022MNRAS.514.3113K}. We may speculate that baryons may be more tightly related to their dark matter halo in the \auriga{} than \eagle{}, and thus we see a stronger correlation between galaxy size and halo spin relation in the \auriga{} simulation than in \eagle{}. Whether this speculation is true and how the detailed feedback physics operates, we will explore in a future study.

\section*{Acknowledgements}
We are grateful to Simon White and Volker Springel for useful discussions. We acknowledge support from NSFC grants (Nos 11988101, 11133003, 11425312, 12033008, 11622325, 11903043), National Key Program for Science and Technology Research Development (2018YFA0404503, 2017YFB0203300), and K. C. Wong Foundation.  LG thanks the hospitality of the Institute for Computational Cosmology, Durham University. CSF acknowledges support by the European Research Council (ERC) through Advanced Investigator DMIDAS (GA 786910). QG acknowledge support from NO. CMS-CSST-2021-A07. RG acknowledges financial support from the Spanish Ministry of Science and Innovation (MICINN) through the Spanish State Research Agency, under the Severo Ochoa Program 2020-2023 (CEX2019-000920-S). SL acknowledges the support by the European Research Council via ERC Consolidator Grant KETJU (No. 818930). QG and SS acknowledge the support from CAS Project for Young Scientists in Basic Research, Grant No. YSBR-062. SS acknowledges the science research grants from the China Manned Space Project with NO. CMS-CSST-2021-B03.

\section*{Data Availability}
The IllustrisTNG simulations, including the TNG100-1 used in this article, are publicly available and accessible at \url{https://www.tng-project.org/data/}. The EAGLE simulation is publicly available at \url{http://eagle.strw.leidenuniv.nl} (see \citet{2016A&C....15...72M} for the original data release description). The other data directly related to this publication and its figures is available on request from the corresponding author.




\bibliographystyle{mnras}
\bibliography{refer} 


\appendix

\section{The dependence of concentration on size--spin relation}
\begin{figure}
	\includegraphics[width=\linewidth]{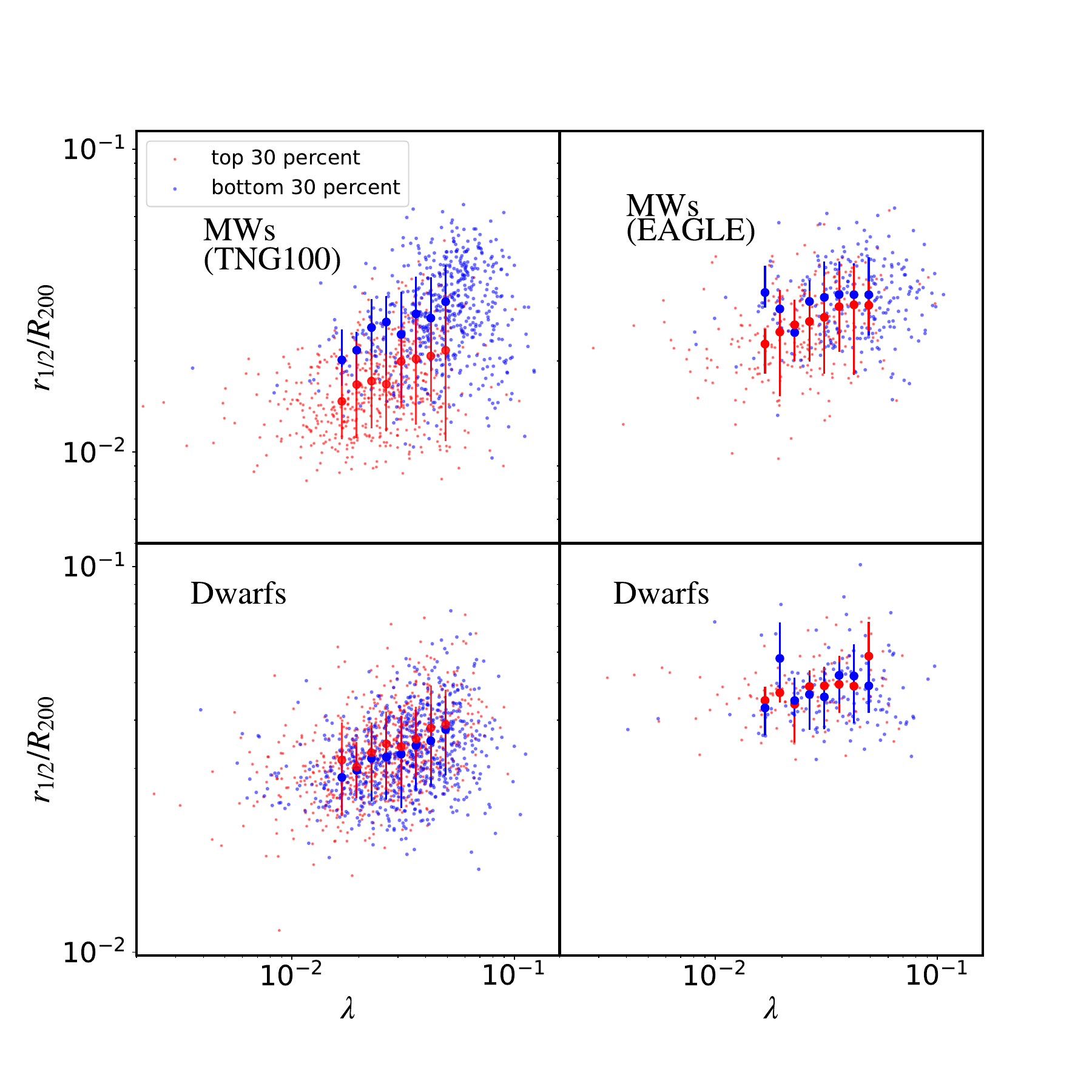}
	\caption{$r_{1/2}/R_{200}$--$\lambda$ relation of MWs (top panels) and disk dwarfs (bottom panels) in IllustrisTNG (left panels) and EAGLE (right panels). The red and blue dots represent large and small concentration samples, respectively. The thick points and error bars show median value and 1$\sigma$ scatter for the corresponding samples.}
    \label{fig:cc}
\end{figure}

We use $c\equiv V_{\rm max}/V_{200}$ to explore the dependence of $r_{1/2}/R_{200}$--$\lambda$ relation on the concentration of the halo. The red (blue) dots in ~\reffig{fig:cc} represent two populations of haloes whose concentration are in the top or bottom 30 per cent of the full sample. For the MW-mass galaxies (top panels), the $r_{1/2}/R_{200}$--$\lambda$ relations between the two subsets are almost identical in the slope but only differ in the amplitude, with the sizes in low-concentrated samples being $1.4 (1.2)$ times larger than those in high-concentrated samples in IllustrisTNG (EAGLE). Thus, for high-mass galaxies, the size of a galaxy relies on both $\lambda$ and concentration. However, for dwarf-mass galaxies (bottom panels), the $r_{1/2}/R_{200}$--$\lambda$ relation between the two samples are very similar, which suggests that for low-mass galaxies the size is mainly determined by $\lambda$.

\section{The dependence of angular momentum retention factor on size--spin relation}

\begin{figure}
	\includegraphics[width=\linewidth]{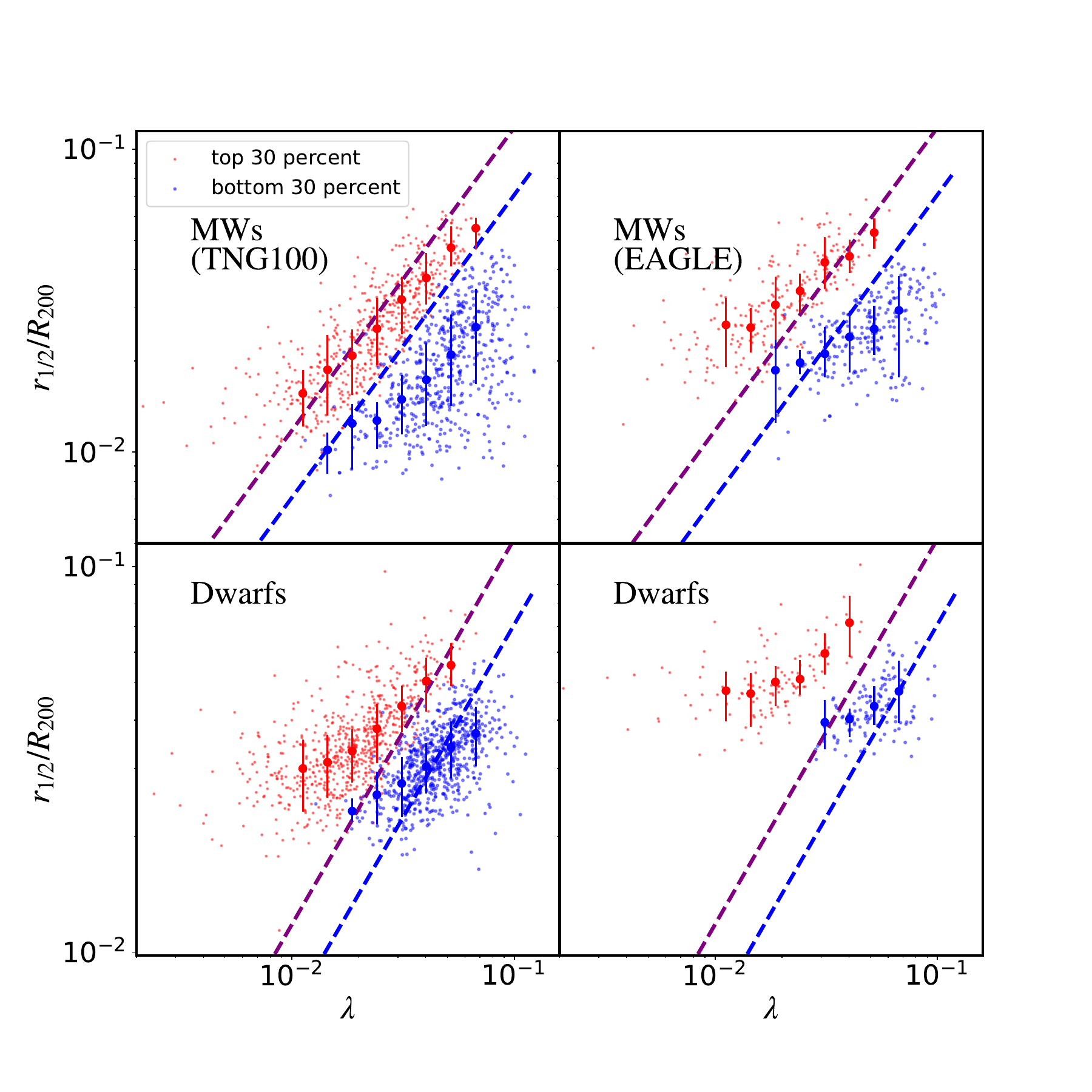}
	\caption{$r_{1/2}/R_{200}$--$\lambda$ relation of MW (top panels) and dwarf analogs (bottom panels) in IllustrisTNG (left panels) and EAGLE (right panels). The red and blue dots represent the top 30 and bottom 30 per cent of the samples according to their angular momentum retention factor $f_j$ samples, respectively. The thick points and error bars show median value and 1$\sigma$ scatter for the corresponding samples. The dashed line display predictions given by the MMW98 model assuming $f_j=1.0$ (purple) and $f_j=0.5$ (blue) with $f_R=1$, respectively.}
    \label{fig:jj}
\end{figure}

\reffig{fig:jj} shows the dependence of the scatter of $r_{1/2}/R_{200}$--$\lambda$ relation on the angular momentum retention factor $f_j=j_{d}/j_{h}$. The red (blue) dots represent the populations whose $f_j$ are in the top (bottom) 30 percent of the full sample. The segregation between the two samples is highly significant, which indicates that $f_j=j_{d}/j_{h}$ is responsible for the scatter in the $r_{1/2}/R_{200}$--$\lambda$ relation.

\bsp	
\label{lastpage}
\end{document}